          \documentclass[manuscript]{aastex}
\usepackage{natbib}
\slugcomment{to appear in ApJ}

\shorttitle{Spectra of Cosmic Ray Protons and Helium Produced in
Supernova Remnants}

\shortauthors{Ptuskin et al.}

\begin{document}

\title{Spectra of Cosmic Ray Protons and Helium Produced in Supernova Remnants}

\author{Vladimir Ptuskin and Vladimir Zirakashvili}

\affil{Pushkov Institute of Terrestrial Magnetism, Ionosphere
 and Radio Wave Propagation of the Russian Academy
 of Science (IZMIRAN), Troitsk, Moscow Region 142190, Russia}

\and

\author{Eun-Suk Seo}

\affil{Department of Physics and Institute of Physical Science and
Technology, University of Maryland, College Park, MD 20742 USA}

\begin{abstract}
Data obtained in the ATIC-2 (Advanced Thin Ionization
Calorimeter), CREAM (Cosmic Ray Energetics and Mass)) and PAMELA
(Payload for Antimatter Matter Exploration and Light-nuclei
Astrophysics) experiments suggest that elemental interstellar
spectra of cosmic rays below the knee at a few times $10^{6}$ GeV
are not simple power laws, but they experience hardening at
magnetic rigidity above about $240$ GV. Another essential feature
is the difference between proton and Helium energy spectra, so
that the He/p ratio increases by more than $50$ \% in the energy
range from $10^{2}$ to $10^{4}$ GV. We consider the concavity of
particle spectrum resulting from the nonlinear nature of diffusive
shock acceleration in supernova remnants (SNR) as a possible
reason for the observed spectrum hardening. Helium-to-proton ratio
increasing with energy can be interpreted as a consequence of
cosmic ray acceleration by forward and reverse shocks in SNRs. The
contribution of particles accelerated by reverse shocks makes the
concavity of the produced overall cosmic ray spectrum more
pronounced. The spectra of protons and helium nuclei accelerated
in SNRs and released into the interstellar medium are calculated.
The derived steady state interstellar spectra are in reasonably
good agreement with observations.

\end{abstract}


\section{Introduction}

High-accuracy measurements have revealed deviations of cosmic ray
spectra from plain power laws at energies $10$ to $10^{5}$ Gev/n.
This refers, in particular, to the ATIC-2 \citep{Panov09}, CREAM
\citep{Ahn10,Yoon11} and PAMELA \citep{PAM} experiments; see
\citet{Lav11} for additional references. The general conclusion
from these measurements is the presence of hardening in the
spectra of protons, helium and probably heavier nuclei at magnetic
rigidity about $240$ GV and the increase of He/p ratio in the
above energy range.

A number of explanations of these results have been suggested: the
hardening may reflect the contribution of two distinct populations
of cosmic ray sources \citep{ZatsSok06}; the action of numerous
sources with the dispersion of source injection spectra
\citep{QYuan11}; fluctuations produced by local cosmic ray sources
\citep{Horandel12}; specific conditions of cosmic ray transport in
the local bubble \citep{ErlWolf12}; interstellar diffusion with a
diffusion coefficient that is not separable in energy and space
\citep{Tomassetti12}; or the combination of diffusion on the
cosmic ray induced turbulence and the background turbulence
\citep{Blasi12}. The energy dependent He/p ratio may be due to
acceleration by shock propagating through the medium with varying
chemical composition, including the helium wind of a Wolf-Rayet
star \citep{PtZirSeo10}, the stratified material of a bubble
\citep{Ohira11}, or rate of ionization \citep{Drury11}, and the
specifics of thermal ion injection in the process of shock
acceleration \citep{Malkov11}. \citet{Bierm} suggested that the
spectral hardening and enrichment in heavy nuclei is due to the
contribution of a polar cap cosmic-ray component produced by
supernova explosions in winds of massive progenitor stars. The
consideration of various interpretations of cosmic ray spectral
peculiarities can be found in the paper \citet{Vlad12}.

In the present work we further develop our model of cosmic ray
acceleration in supernova remnants \citep{PtZirSeo10} (Paper I) to
explain both of the required features. To a good approximation,
the model in Paper I explains the observed overall spectrum of
cosmic rays in a wide range of energies up to about $10^{9}$ GeV.
An important improvement of the code used in Paper I was made in
our work \citep{ZirPt11} (Paper II), where particle acceleration
by a reverse (backward) shock moving through the material of
supernova ejecta was included in the calculations. This is in
addition to the acceleration by the forward shock moving through
the circumstellar medium studied in Paper I. The code described in
Paper II was used by \citet{ZirAhar,ZirAhar11} for modelling
particle acceleration in SNRs RXJ1713.7-3946 and Cas A.

It is worth noting that similar numerical models of nonlinear
production of cosmic rays in SNRs, where time-dependent cosmic-ray
transport equations are solved together with gas-dynamic equations
in sperical symmetry were developed by two other groups of authors
\citep{Berezhko96,BV} and \citep{TJ}. They give results analogous
to our model for particle acceleration by forward SNR shocks, but
they do not include acceleration by reverse shocks. Cosmic-ray
acceleration by reverse SNR shocks was earlier considered by
\citet{BerPt}. The process was recently analysed by \citet{Telezh}
in a test particle approximation.

Two effects considered below in the frameworks of our model of
particle acceleration in SNRs are of particular importance: the
nonlinear shock modification that leads to the concave spectrum of
accelerated particles with a pronounced hardening; and taken into
account acceleration of supernova ejecta material poor in hydrogen
by a reverse shock that leads to the difference between the
overall proton and helium spectra and increases their concavity.

\section{Modelling of cosmic-ray acceleration in supernova remnants}

Cosmic ray acceleration in shell SNRs proceeds through the
diffusive shock acceleration mechanism that is a version of the
first order Fermi acceleration in the shock vicinity, where the
gas with a frozen magnetic field is compressing. The fast, charged
background particles are scattered by random magnetohydrodynamic
(MHD) waves and inhomogeneities, and they gain energy crossing the
shock, see e.g. \cite{MalkDrury01} for a review. The process of
efficient acceleration should be modelled simultaneously with the
SNR hydrodynamics, because the pressure of accelerated particles
modifies the gas flow and the shock structure that affects the
particle spectrum. Also, the current of accelerated particles
 results in the streaming instability that
amplifies the background MHD waves. These waves in turn
determine the value of particle spatial diffusion.

Hydrodynamical equations for the gas density  $\rho (r,t)$, gas
velocity $u(r,t)$, gas pressure $P_g(r,t)$, and the equation for
the isotropic part of the cosmic-ray proton momentum distribution
 $N(r,t,p)$ in the spherically symmetrical case take the form
 (see also Paper II):

\begin{equation}
\frac {\partial \rho }{\partial t}=-\frac {1}{r^2}\frac {\partial
}{\partial r}r^2u\rho,
\end{equation}

\begin{equation}
\frac {\partial u}{\partial t}=-u\frac {\partial u}{\partial
r}-\frac {1}{\rho } \left( \frac {\partial P_g}{\partial r}+\frac
{\partial P_c}{\partial r}\right),
\end{equation}

\begin{equation}
\frac {\partial P_g}{\partial t}=-u\frac {\partial P_g}{\partial
r} -\frac {\gamma _gP_g}{r^2}\frac {\partial r^2u}{\partial r}
-(\gamma _g-1)(w-u)\frac {\partial P_c}{\partial r},
\end{equation}

\[
\frac {\partial N}{\partial t}=\frac {1}{r^2}\frac {\partial
}{\partial r}r^2D(p,r,t) \frac {\partial N}{\partial r} -w\frac
{\partial N}{\partial r}+\frac {\partial N}{\partial p} \frac
{p}{3r^2}\frac {\partial r^2w}{\partial r}
\]
\[
+\frac {\eta ^f\delta (p-p_{f})}{4\pi p^2_{f}m}\rho
(R_f+0,t)(\dot{R}_f-u(R_f+0,t))\delta (r-R_f(t))
\]
\begin{equation}
+\frac {\eta ^b\delta (p-p_{b})}{4\pi p^2_{b}m}\rho
(R_b-0,t)(u(R_b-0,t)-\dot{R}_b)\delta (r-R_b(t)).
\end{equation}

Here $P_c=4\pi \int p^2dpvpN/3$ is the cosmic-ray pressure,
$w(r,t)$ is the advective velocity of cosmic rays, $\gamma _g$ is
the adiabatic index of the gas, and $D(r,t,p)$ is the cosmic-ray
diffusion coefficient. It was assumed that the diffusive streaming
of cosmic rays results in the generation of MHD waves. Cosmic-ray
particles are scattered by these waves. That is why the cosmic-ray
advective velocity $w$ may differ from the gas velocity $u$. In
our modelling they differ on the value of the radial component of
the Alfv\'en velocity calculated in the isotropic random magnetic
field: $w=u+\xi _AV_{A}/\sqrt{3}$, $V_A=B/\sqrt{4\pi \rho }$ .
Here the factor $\xi _A$ describes the possible deviation of the
cosmic ray drift velocity from the gas velocity. The values $\xi
_A=1$ and $\xi _A=-1$ are assumed upstream of the forward and
reverse shocks respectively, where Alfv\'en waves are generated by
the cosmic ray streaming instability and propagate in the
corresponding directions. The damping of these waves heats the gas
upstream of the shocks, \cite{mckenzie82}, which is described by
the last term in Eq. (3) and limits the total compression ratios
by a number close to 6. In addition, we use the value $\xi _A=-1$
downstream of the shocks because the cosmic ray gradient is
positive in this region. The Alfven drift here strongly influences
the slope of the particle momentum spectrum, since magnetic fields
are compressed downstream of the shocks, and the gas velocity in
the shock frame falls below the sound speed. This leads, in
particular, to noticeable steepening of the particle spectrum at
the forward shock, see Paper I.

The two last terms in Eq. (4) correspond to the injection of
thermal protons with momenta $p=p_{f}$, $p=p_{b}$ and mass $m$ at
the fronts of the forward and reverse shocks at $r=R_f(t)$ and
$r=R_b(t)$, respectively. The indexes $f$ and $b$ are used for
quantities corresponding to the forward and reverse shock
respectively. The dimensionless parameters $\eta ^f$ and $\eta ^b$
determine the injection efficiency. The magnetic energy density is
relatively small, so it does not appear explicitly in Eq. (2).

Shocked ejecta and interstellar gas are separated by the contact
discontinuity at $r=R_c $. The spatial dependence of the magnetic
field at $r>R_c$ is taken in the form
\begin{equation}
B(r,t)=\sqrt{\frac{4\pi}{\rho_{0}}}
\frac{\rho\dot{R_{f}}}{M_{A}}\sqrt{1+(M_{A}V_{A0}/\dot{R_{f}})^{2}},
 \ r>R_{c}.,
\end{equation}
where $\rho _0$ is the gas density of the circumstellar medium,
$V_{A0}$ is the Alfv\'en velocity there, and $M_A$ is some
constant. We employ the results of \citet{Voelk05} in analysis of
X-ray radiation from young SNRs, and we assume that magnetic
energy density $B^{2}/8\mathrm{{\pi}}$ downstream of the shock is
$3.5$\% of the ram pressure $\rho _0\dot{R_{f}}^{2}$ that
determines the constant $M_A=23$. According to Eq.(5) the far
upstream energy density of the  amplified magnetic field is a
small $ 0.5/M_A^2$ part of the ram pressure. Note that this
relation is consistent with modelling of the cosmic ray streaming
instability in young SNRs \citep{Zir08b}. We assume that the
magnetic field is compressed in accordance with plasma density
upstream and downstream of the forward shock. We also assume that
there is no
 decay of the magnetic field in the downstream region.
 The magnetic amplification is weak in old remnants with the forward shock
 speed $\dot{R_f}<M_AV_{A0}$. The magnetic field strength $B$ is
close to the interstellar value $B_0$ in these SNRs (see Eq. (5)).
We assume that the magnetic field does not depend on radius
downstream of the reverse shock at $R_b<r<R_c$, but depends on
{\bf density} similarly to Eq. (5) upstream of the reverse shock,
$r<R_b$.

The diffusion coefficient is of the form
\begin{equation}
D=\kappa D_{B}.
\end{equation}
Here $D_{B}=\mathit{vpc}/(3\mathit{ZeB})$ is the so called Bohm
diffusion coefficient for particles of charge $\mathit{Ze}$ and
velocity $v$ ($c$ is the speed of light). The function
$\kappa=(1+(M_{A}V_{A0}/\dot{R_f})^2)^{3}$ at $r>R_{c}$ and
$\kappa=3$ at $r<R_{c}$ approximates the dependence of the
diffusion coefficient on shock velocity. It corresponds to the
case when the MHD turbulence in the shock vicinity is amplified by
cosmic ray streaming instability, which is balanced by the
Kolmogorov-type nonlinearity, see \cite{ptuskin05} and Paper II
for details. We assume that the diffusion coefficient is close to
its Bohm limit in young SNRs when the shock velocity exceeds $\sim
700$ km/s. This slightly underestimates the diffusion coefficient
of particles
 at the very end of the energy spectrum. This is
 because these particles are scattered by small-scale
magnetic inhomogenuities generated by the non-resonant streaming instability
and their diffusion coefficient
is several times higher than the Bohm diffusion
 coefficient \citep{Zir08b}.
The
value of $D$ grows with time because the wave generation is
becoming less efficient as the shock velocity is decreasing.

The equation for ions is similar to Eq. (4). For ions with mass
number $A$ and mass $M=Am$, it  is convenient to use the momentum
per nucleon $p$ and the normalization of the ion spectra $N_i$ to
the nucleon number density. Then, the number density of ions $n_i$
is  $n_i=4\pi A^{-1}\int p^2dpN_i$. The ion pressure $P_i=4\pi
\int p^2dpvpN_i/3$ is also taken  into account in the total cosmic
ray pressure $P_c$.

The numerical procedure we use to solve Eqs. (1)-(4) was described
in detail in Paper II. A finite-difference method is employed to
Eqs. (1)-(4) upstream and downstream of the forward and reverse
shocks. A non-uniform numerical grid upstream of the shocks at
 $r>R_f$ and $r<R_b$ allows to resolve small scales of hydrodynamical
quantities resulting from the pressure gradient of low-energy
cosmic rays. The gases compressed at the forward and reverse
shocks are separated by a contact discontinuity at $r=R_c$ between
the shocks. An explicit conservative TVD scheme for hydrodynamical
equations (1)-(3) and uniform spatial grid are used between the
shocks.

\section{Spectra of cosmic rays produced by supernova remnants}

To demonstrate a possible effect of acceleration by reverse shocks
on cosmic-ray composition, we make calculations for two classes of
supernovae: Type I where Hydrogen is absent in the outer layers of
ejecta and Type II where Hydrogen strongly prevails. The relative
rates of these types of supernovae in the Galaxy are $0.46$ and
$0.54$ respectively \citep{Fillip}. The following characteristics
of Type Ia and Type IIP SNRs as representative of these two
classes of supernovae are accepted in the calculations.

Type Ia SNRs have kinetic energy of explosion
$E={10}^{51}\mathrm{erg}$, number density of the surrounding
interstellar gas $n=0.1$ cm${}^{-3}$, and mass of ejecta
$M_{\mathrm{ej}}=1.4M_{\odot }$. Also important for accurate
calculations is the index $k$, which describes the power law
density profile $\rho _{s}\propto r^{-k}$ of the outer part of the
star that freely expands after supernova explosions; $k=7$ for
Type Ia supernova. Type IIP SNRs have parameters $E$ = ${10}^{51
}\mathrm{erg}$, $n=0.1 $ ${\mathrm{cm}}^{-3}$,
$M_{\mathrm{ej}}=8M_{\odot }$, and $k=12.$

\begin{figure}[!t]
  \vspace{5mm}
  \centering
   \includegraphics[width=7.0 cm]{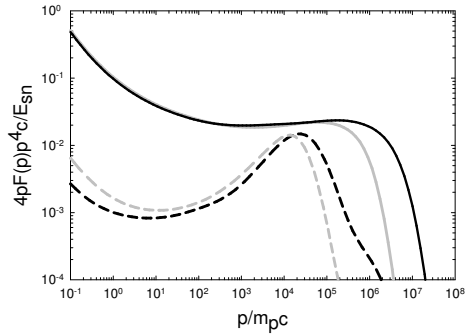}
  \caption{Spectra of nucleons produced during SNR lifetime by forward (solid lines)
   and reverse shocks (dash lines) in Type I (black lines)
   and Type II (gray lines) SNRs as functions of momentum
   per nucleon $p$. The case is shown when only protons are accelerated
   by forward shocks in the interstellar medium and by the reverse shocks in
   Type II SNRs whereas only nuclei with mass to charge ratio $A/Z=2$ are
   accelerated by the reverse shocks in Type I SNRs.}
  \label{ZirSource}
 \end{figure}

Fig.\ref{ZirSource} shows the overall spectra of relativistic
nuclei produced in Type Ia and Type IIP SNRs during $10^{5}$ years
of their evolution by forward (solid lines) and reverse (dash
lines) shocks. It is assumed for illustration that the
interstellar medium and the ejecta of Type II SNRs consist of
protons; the ejecta of Type I SNRs consist of nuclei with the mass
number $A$ ($A>1$) and the charge $Z=A/2$. The accepted value of
the interstellar magnetic field is $5 \mu$G. The pressure of
accelerated particles modify the profiles of forward and reverse
shocks that leads to the concave spectra of cosmic rays. The
plotted function is defined as $Q(p)=4\pi p^{2}F(p)$, where $F(p)$
is the distribution of all accelerated nucleons injected in the
interstellar medium over the SNR lifetime. The total number of
accelerated nucleons is $\int Q(p) dp$. The procedure for
calculating SNR evolution and simultaneous cosmic-ray acceleration
was described in-depth in Paper II. One can also find there the
plots that show the calculated profiles of the gas density,
velocity and pressure, and the cosmic ray pressure in evolving
SNRs.

The spectra of particles accelerated at reverse shocks are harder
than the spectra at forward shocks, in spite of the same level of
the shock modification for both shocks. This is because the shocks
propagate in the media with different properties. When the reverse
shock reaches the flat part of the ejecta density distribution it
propagates in the medium density which is decreasing with time.
That is why the number of freshly injected ions is low in
comparison with higher energy particles accelerated earlier,
thereby leaving to spectral hardening. This effect is absent at
the forward shock propagating in the medium with constant density.

Particles accelerated in the numerous SNRs are injected into
interstellar space, diffuse in galactic magnetic fields, interact
with interstellar gas, and finally escape through the cosmic ray
halo boundaries into intergalactic space, where cosmic ray density
is negligible. We employ the plain diffusion model with a flat
cosmic ray halo \citep{GP76, StrongRev} for calculations of cosmic
ray propagation in the Galaxy. The leaky-box approximation to the
diffusion model can be used for our purpose - the determination of
proton and helium intensities, see \citet{PtSS}. The cosmic ray
intensity obeys the relation $I\propto\nu
_{\mathrm{sn}}Q(X_{e}^{-1}+\sigma/m_{a})^{-1}$, where $X_{e}$ is
the escape length (the average matter thickness traversed by
cosmic rays before they exit from the Galaxy), $\sigma$ is the
nuclear spallation cross section for a given type of relativistic
nuclei moving through the interstellar gas, and $m_{a}$ is the
mean interstellar atom mass. The escape length is determined from
the relative abundance of secondary nuclei (primarily from the
Boron-to-Carbon ratio) in cosmic rays. The approximation formula
$X_{e}=19\beta^{3}$ g/cm$^{2}$ at $R \leq 3$ GV and
$X_{e}=19\beta^{3}(R/3 \textrm{GV})^{-0.6}$ g/cm$^{2}$ at $R > 3$
GV was given in \citet{PtSS} (here $R$ is the particle magnetic
rigidity; $\beta=v/c$). According to the last equation, the
resulting spectrum is steeper than the source spectrum by ${0.6}$
at high enough energies, but the uncertainty in the last value is
about $0.1$, and statistically accurate measurements of the
Boron-to-Carbon ratio are not available at energies above $\sim30$
GeV/nucleon. In the present calculations we assume  the dependence
$X_{e}\propto R^{-0.7}$ at $R>3$ GV up to rigidity $\sim 10^{6} $
GV. It corresponds to dependence of the cosmic-ray diffusion
coefficient on rigidity in the interstellar medium $D_{ISM}\propto
R^{0.7}$ at relativistic energies.

\begin{figure}[!t]
  \vspace{5mm}
  \centering
  \includegraphics[width=10.0 cm]{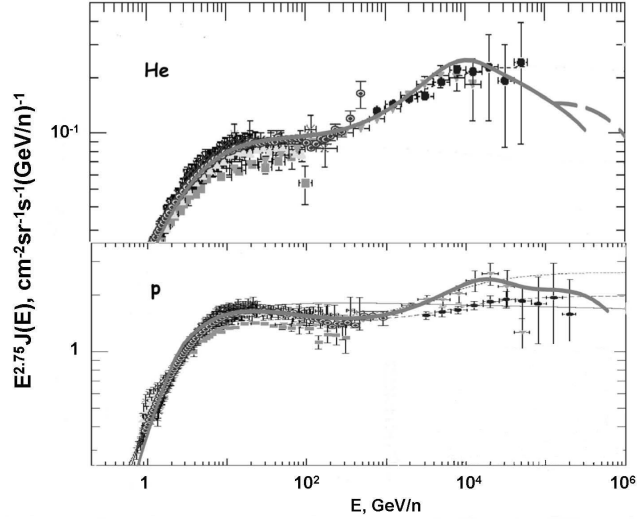}
  \caption{Calculated interstellar spectra of protons and helium
  are shown by thick gray lines. Data of the BESS, CAPRICE, AMS-01,
  ATIC-2 and CREAM experiments corrected for solar modulation effects
   are taken from the review of
  \citet{Lav11}, where the corresponding references and
  detailed description of observations can be found.
  The PAMELA experiment data are taken from the paper \citet{PAM}
  and corrected for solar modulation.}
  \label{twoShocks}
 \end{figure}

The interstellar proton and helium cosmic-ray spectra determined
in the framework of this propagation model, and use the source
functions calculated as described above, are presented in
Fig.\ref{twoShocks}. It is assumed that the reverse shocks
accelerate only hydrogen nuclei in Type II SNRs and only helium
nuclei in $65$\% of all Type I SNRs; the reminding $35$\% are Type
Ic SNRs, which do not have significant amounts of hydrogen and
helium in their ejecta. The forward shocks produced by Type I and
Type II supernovae propagating through the interstellar gas of
standard composition with the relative injection efficiency of
hydrogen and helium ions at the forward shocks is chosen to fit
the observed He/p ratio at the $100$ GeV/nucleon reference energy.
The shock acceleration of background particles with different
elemental composition leads to the difference in the overall
proton and helium spectra seen in Fig.\ref{twoShocks}. The forward
shock in Type Ib/c SNRs ($59$\% of all Type I SNRs) moves through
the helium wind of the progenitor Wolf-Rayet star, the last stage
of a progenitor star evolution for these supernovae, that enriches
cosmic rays in helium at energies above approximately $10^{5}$
GeV/n. The possible contribution of this process is shown by dash
line in Fig.\ref{twoShocks}.

The calculated spectra in Fig.\ref{twoShocks} exhibit hardening at
$\sim 240$ GV and reproduce the energy dependent He-to-proton
ratio in fair agreement with observations. Thus, the explanation
of these features in terms of acceleration of cosmic rays in SNRs
by forward and reverse modified shocks seems quite possible.

\section{Discussion and conclusion}
The processes of cosmic-ray acceleration by supernova shocks and
subsequent diffusion in the interstellar magnetic fields at
relativistic energy depend on particle Larmor radius so the
spectra of different ions might be expected to have similar shapes
when expressed as functions of particle magnetic rigidity.
However, the data of ATIC-2 \citep{Panov09}, CREAM
\citep{Ahn10,Yoon11} and PAMELA \citep{PAM} experiments showed
that the He/p ratio is changing with energy at $10$ to $10^{5}$
GeV/n. Also, the spectra of these ions demonstrate deviations from
the plain power laws with hardening at about $240$ GV. Thus,
modern statistically accurate measurements confront the
traditional power-law paradigm for the source spectrum and escape
length of Galactic cosmic rays below the knee. A number of
possible explanations of these experimental results were proposed
in the papers cited in the Introduction, but a clear picture does
not yet emerge.

In the present paper, the interpretation of both features
(spectral hardening and changing of the He/p ratio) is provided in
the framework of our model of cosmic ray acceleration by SNR
shocks developed in Papers I and II (the first version of this
work was presented in the conference paper \citet{Beijing}). The
hardening of cosmic ray spectrum results mainly from modification
of gas flow in the shock precursor by the cosmic ray pressure,
which shapes the concave energy spectrum of cosmic rays. This
effect is well known in the theory of diffusive shock
acceleration, see e.g. early papers \citet{Eichler, Berezhko96}.
It manifests itself here in the presence of Alfven drift of
accelerated particles and the strong dependence of cosmic-ray
diffusion on shock velocity as a pronounced hardening at rigidity
$\sim240$ GV in the overall spectrum of accelerated particles
produced by SNRs. The account of particle acceleration by a
reverse shock moving through the supernova ejecta results in the
production of an additional component of cosmic rays that has a
specific hard energy spectrum and is depleted in hydrogen
composition. This makes hardening of the overall cosmic ray
spectrum more pronounced, and it produces the difference between
hydrogen and helium spectra.

The results of our calculations of the interstellar spectra of
cosmic-ray hydrogen and helium shown in Fig.~\ref{twoShocks}
demonstrate good agreement with observations. It is assumed that
reverse shocks accelerate only protons in Type II SNRs and only
helium nuclei in Type I SNRs. Forward shocks accelerate
interstellar protons and helium nuclei in both cases.

Clearly more work is needed and planned to establish the adequacy
of the suggested scenario. This refers both to the detailed
analysis of the dispersion properties of SNRs (the SN explosion
energy, the mass and complex composition of ejecta, the structure
of circumstellar medium, etc.) and assumptions about the process
of particle acceleration (the value of $M_{A}$, the
parametrization of diffusion coefficient, etc.). Also, a single
power-law dependence of cosmic-ray diffusion in the interstellar
magnetic fields on rigidity, $D_{ISM}\propto vR^{0.7}$, used in
the present calculations in the entire range of rigidities from
$10$ to $10^{6}$ GV is most probably a significant
oversimplification. The explanation of low anisotropy of Galactic
cosmic rays seems incompatible with such a strong dependence of
diffusion on rigidity \citep{anis,AmatoBlasi}.

In principle, the last difficulty could be relieved by increasing
$M_{A}$ from its $M_{A}=23$ value taken from Paper I and accepted
in the present calculations. This would lead to a steeper source
spectrum and require weaker dependence of interstellar diffusion
on rigidity to fit the observations of cosmic ray spectra. Another
possibility is a possible suppression of cosmic-ray anisotropy in
the Local Bubble with enhanced turbulence where the Solar system
resides, \cite{zirakashvili05}.

\acknowledgments This work was supported by by the NASA Astronomy
and Physics Research and Analysis grant NNX09AC14G and the Russian
Foundation for Basic Research grant 10-02-00110a.

\bibliography{hardening}

\end{document}